\documentclass[final]{parallel}
\usepackage{bm}
\usepackage{oldlfont,alltt,amsmath}
\usepackage{epsfig}
\usepackage{graphicx}
\usepackage{makeidx}
\usepackage{multicol}
\usepackage{subeqn}
\usepackage{crop}
\crop
\newcommand{\vertexset}{V}
\newcommand{\vertexsubset}{V}
\newcommand{\edgeset}{E}
\newcommand{\edgsubeset}{E}
\newcommand{\convbody}{B}
\newcommand{\ppp}{P}
\newcommand{\vandermond}{W}

\newcommand{\cachec}{S}
\newcommand{\cachea}{a}
\newcommand{\cachel}{z}
\newcommand{\cacheb}{w}
\newcommand{\memadr}{A}

\newcommand{\grpppp}{U}
\newcommand{\numgp}{|V|}
\newcommand{\grdim}{d}

\newcommand{\bdry}{B}

\newcommand{\const}{c}
\newcommand{\uniformg}{starry}
\newcommand{\Uniformg}{Starry}
\newcommand{\hcutth}{Hyperplane Cut Theorem}

\newcommand{\expop}{K}
\newcommand{\expopr}{r}

\newcommand{\cuttree}{T}
\newcommand{\dmood}{\frac{\grdim-1}{\grdim}}
\newcommand{\dmin}{{-\frac{1}{\grdim}}}
\newcommand{\ugrid}{$G=(\vertexset,E)$ }

\begin{document}
%\chapter{Using Minimum-Surface Bodies for Iteration Space Partitioning}
%\chapter{Minimizing Cache Misses Using Minimum-Surface Bodies}
\chapter{Minimizing Cache Misses in Scientific Computing Using
Isoperimetric Bodies}
\begin{authorline}
Michael Frumkin and Rob F.\ Van der Wijngaart\thanks{Computer
        Sciences Corporation; M/S T27A-2, NASA Ames Research Center,
        Moffett Field, CA 94035-1000; e-mail:
        \texttt{\{frumkin,wijngaar\}@nas.nasa.gov}
        NASA Advanced Supercomputing Division
        NASA Ames Research Center}
\end{authorline}

A number of known techniques for improving cache performance in
scientific computations involve the reordering of the iteration space.
Some of these reorderings can be considered coverings of the iteration
space with sets having small surface-to-volume ratios.  Use of such
sets may reduce the number of cache misses in computations of local
operators having the iteration space as their domain.  First, we
derive lower bounds on cache misses that any algorithm must suffer
while computing a local operator on a grid.  Then, we explore
coverings of iteration spaces of structured and unstructured
discretization grid operators which allow us to approach these lower
bounds.  For structured grids we introduce a covering by successive
minima tiles based on the interference lattice of the grid.  We show
that the covering has a small surface-to-volume ratio and present a
computer experiment showing actual reduction of the cache misses
achieved by using these tiles.  For planar unstructured grids we show
existence of a covering which reduces the number of cache misses to
the level of that of structured grids.  Next, we introduce a class of
multidimensional grids, called \uniformg\ grids in this paper.  These
grids represent an abstraction of unstructured grids used in, for
example, molecular simulations and the solution of partial
differential equations.  We show that \uniformg\ grids can be covered by
sets having a low surface-to-volume ratio and, hence have the same
cache efficiency as structured grids.  Finally, we present a
triangulation of a three-dimensional cube that has the property that
any local operator on the corresponding grid must incur a
significantly larger number of cache misses than a similar operator on
a structured grid of the same size.

\section{Introduction}\label{sec:INTRODUCTION}
  A number of known techniques for improving cache performance in
scientific computations involve the reordering of the iteration
space\footnote{In \cite{FRUMKIN_WIJNGAART} we laid the foundation for
the study of reorderings to improve cache performance of local
operators on structured grids. 
 We extend those results in this paper
to the construction of practical, near-optimal tilings of structured
grids, and to the study of cache misses for local operators on
unstructured grids.}.  
We present two new methods for partitioning structured grid iteration
spaces with minimum-surface cache fitting sets. 
 Such
partitionings reduce the number of cache misses to a level that is
close to the theoretical minimum. 
We demonstrate this reduction by
actual measurements of cache misses in computations of a second order
stencil operator on structured three-dimensional grids. 

 A good tiling
of the iteration space for structured discretization grids in the
presence of limited cache associativity can be constructed by using
the interference lattice of the grid. 
 This lattice is a set of grid
indices mapped into the same word in the cache or, equivalently, a set
of solutions of the Cache Miss Equation \cite{MARTONOSI}. 
 In
\cite{FRUMKIN_WIJNGAART} we introduced a (generally skewed) tiling of
the iteration space of stencil operators on structured grids with
parallelepipeds spanned by a reduced basis of the interference lattice.
We showed that for lattices whose second shortest vector is relatively
long, the tiling reduces the number of cache misses to a value close to
the theoretical lower bound. 
 Constructing the skewed tiling, however,
is a nontrivial task, and involves a significant overhead in testing
whether a particular point lies inside the tile. 
 Tiling a
three-dimensional grid, for example, requires the determination of 29
integer parameters to construct the loop nest of depth six, and
involves a significant branching overhead. 

 We start the paper with
deriving lower bounds on the number of cache misses that any algorithm must suffer
while computing a local operator on either structured or
unstructured grids\footnote{The derivation is a simplification of the
proof given in \cite{FRUMKIN_WIJNGAART} for structured grids and
extends its application to unstructured grids.}.  
Subsequently, we
introduce two new coverings of structured grids: Voronoi cells and
rectilinear parallelepipeds built on the vectors of successive minima
of the grid interference lattice. 
 In lattices with a relatively long
shortest vector the cells of both coverings have near-minimal
surface-to-volume ratios. 
 Hence, the number of cache misses in the
computations tiled with these cells is close to the theoretical
minimum derived in \cite{FRUMKIN_WIJNGAART}. 

 For the computation of local discretization operators on planar, fixed-degree 
unstructured grids we construct a
near-minimum-perimeter covering by applying the Lipton-Tarjan planar
graph separator algorithm \cite{LIPTON}. 
 The perimeter-to-area ratio
of the sets of this covering is ${\cal O}(1/\sqrt \cachec)$, where $\cachec$ 
is the cache size. 
Next, we introduce a class of multidimensional grids, called 
\uniformg\ grids,
and show that a $\grdim$-dimensional \uniformg\ grid can be covered by 
sets having a surface-to-volume ratio close to that of structured grids, i.e. 
${\cal O}(\cachec ^{-1/\grdim})$.

Lastly, we construct a fixed-degree unstructured grid that triangulates a
three-dimensional cube and show that the grid can not be covered with sets
having small surface-to-volume ratios. 
 The last result shows that any
computation of a local operator on such a three-dimensional grid must
suffer a larger number of cache misses than the computation of a
similar operator on a structured grid of the same size. 
 Hence,
general unstructured grids of dimension higher than two are provably
less efficient in a cache utilization than structured grids.

\section{Cache and Discrete Geometry}\label{sec:CACHE_USAGE}

\textit{Local operators on grids.} 
We consider the problem of computing array $q$ by applying a local 
discretization operator $\expop$ to array $u$ (i.e.\ $q=\expop u$) defined 
at the vertices of an undirected graph \ugrid\footnote{We are particularly 
interested in graphs that can be interpreted as discretization  grids, and 
will often use the words ``grid'' and ``graph'' interchangeably.}. 
Locality of $\expop$ means that computation of $q(x),~x \in V$, 
involves values of $u(y),~y \in V$, where $y$ is at a (graph) 
distance\footnote{The graph distance is the length of a shortest path 
connecting two vertices. 
The length of the path is the number of edges it contains.} at most $ \expopr $ from $x$. 
This $\expopr$ is called the order of $\expop$ and is assumed to be independent of $G$. 
Arrays $q$ and $u$ are distinct.
Hence, the values of $q$ can be computed in arbitrary order.

We consider structured and unstructured grids that have an explicit or implicit 
embedding into Euclidean space. 
Structured grids can be defined as Cartesian products of line graphs, while edges of 
unstructured  grids are defined explicitly, for example by an adjacency matrix. 
We assume that the maximum vertex degree is independent of the total number of vertices. 
A grid is called planar if its vertices and edges can be embedded into a plane without 
edge intersections. 
A grid is called a triangulation of a body $\convbody$ if it can be represented 
as a one-dimensional skeleton of a simplicial partition of $\convbody$ 
(see \cite{EDELSBRUNNER}). 

\textit{Cache Model.} 
We consider a single-level, virtual-address-mapped, set-as\-socia\-tive data cache
memory, see \cite{HENNESSY}.
The memory, with a total capacity of $\cachec$ words, is organized in $\cachel$ 
sets of $\cachea$ (\textit{associativity}) lines each.
Each line contains $\cacheb$ words.
Hence, the cache can be characterized by the parameter triplet $(\cachea,\cacheb,\cachec)$, 
and its size $\cachec$ equals $\cachea*\cachel*\cacheb$ words.
A cache with parameters $(\cachec/\cacheb,\cacheb,\cachec)$ is called fully associative, 
and a cache with parameters $(1,\cacheb,\cachec)$ is called direct-mapped.

The cache memory is used as a temporary fast storage of words used for
processing.
A word at virtual address $A$ is fetched into  cache location 
$(\cachea(\memadr),\cachel(\memadr ),\cacheb(\memadr ))$, where $\cacheb(\memadr ) = \memadr \bmod \cacheb$, $\cachel (\memadr) = (\memadr /\cacheb) \bmod \cachel$, and $\cachea(\memadr)$
is determined according to a replacement policy (usually a variation of
\textit{least recently used}).
The replacement policy is not important within the scope of this paper since our lower 
bounds are valid for any replacement policy and our upper bounds hold even for a 
direct mapped cache.

The number of cache misses incurred in computation of $q$ depends on the order in which elements of $u$ are stored in the main memory. 
We assume that for structured grids an element $u(i_1,\dots,i_\grdim)$ is stored at address $\memadr=i_1 + n_1 i_2 + n_1 n_2 i_3 + \cdots + n_1 \cdots n_{\grdim-1} i_\grdim$, where $n_1, \cdots , n_{\grdim-1}$ are the grid sizes. 
For unstructured grids we don't assume any particular ordering in memory 
of the grid points (and, hence, of elements of $u$). 
Instead, we choose an ordering that reduces the number of cache misses.

\textit{Replacement loads.} A \textit{cache miss} is defined as a request for a word of data that is not present in the cache at the time of the request. 
A \textit{cache load} is defined as an explicit request for a word of data for which no explicit request has been made previously (a \textit{cold load}), or whose residence in the cache has expired because of a cache load of another word of data into the exact same location in the cache (a \textit{replacement load}). Cache load is used as a technical term for making some formulas and their proofs shorter. 
The definitions of cold and replacement loads are analogous to those of cold and replacement cache misses \cite{MARTONOSI}, respectively, and if $\cacheb$ equals 1 they completely coincide.

\textit{Isoperimetric Sets}. 
In addition to inevitable cold loads, some replacement loads must occur, as 
formalized in Lemma \ref{lemma:LOWER_BOUND_BASE}, because some values have to 
be dropped from cache before they are fully utilized.
This leads us to a lower bound on cache misses, see Theorems 
\ref{th:LOWER_BOUND_POINTWISE}, \ref{th:LOWER_BOUND_SEPARABLE}. 
This bound includes the surface-to-volume ratio of isoperimetric subsets of the 
grid, that is, sets whose boundary size is $3\cachec$ and that have maximum volume.

\textit{Surface-to-volume ratio}. 
Some replacement loads may be avoided by careful scheduling of the computations. 
One technique for minimizing the number of replacement loads is to cover the 
grid $G$ with conflict-free sets $\vertexsubset_i$ 
($\vertexset=\bigcup_{i=1}^k \vertexsubset_i, |\vertexsubset_i|=\cachec$), 
that is, sets whose vertices are all mapped to different locations in cache. 
If we calculate $q$ in all vertices of $\vertexset_i$ before calculating it in 
vertices of $\vertexsubset_{j},~j=i+1,\dots,k$, a replacement load can occur 
only at vertices having neighbors in at least two sets (boundary vertices). 
We consider only bounded-degree graphs, so if we can find a covering with 
sets $\vertexsubset_i$  having volumes $|\vertexsubset_i|$ close to $\cachec$ 
and a  minimal number of boundary vertices $|\partial \vertexsubset_i|$ 
(and boundary  edges) i.e., bodies with minimum surface-to-volume ratio, then the 
computation of $q$ will incur a number of replacement loads close to the minimum. 
On the other hand, the total partition boundary 
$\sum_{i=1}^{k}|\partial \vertexset_i|$ can be used to obtain a lower 
bound on the number of replacement loads, see section \ref{sec:LOWER}, 
cf.\ \cite{FRUMKIN_WIJNGAART,PEBBLE}.
Deviating from the literal meaning of the word ``isoperimetric'' and following 
common practice, we call sets having the minimum number of boundary points for a 
given number of total points also isoperimetric.

\section{A lower bound on cache misses for local operators}\label{sec:LOWER}
In this section we consider the following problem:
for a given grid $G$ and a local operator $\expop$, how many cache misses 
must be incurred in order to compute $q = \expop u$, where $q$ and $u$ are 
two distinct arrays defined on the grid.
We provide a lower bound on the number of cache misses in any algorithm, 
regardless of the order in which the grid points are visited during the computation 
of $q$.
The lower bound contains the surface-to-volume ratio of isoperimetric sets of the 
grid. 
This ratio can be well estimated for structured grids, and for FFT grids, to
be described in Section \ref{sec:CACHE_UNFRIENDLY}.
It may be shown that our lower bound is tight for the above-mentioned grids,
and in general for any grid which may be covered by sets of size $\cachec$
having an optimal surface-to-volume ratio. 

We use the following terminology to describe the operator $\expop$.
Locality of $\expop$ means that the value of $q$ at grid point $x$ 
is a function of the values $u(y)$, where $y$ is a grid 
point at a graph distance at most $\expopr$ from $x$ ($\expopr$ is 
called the radius, or order, of $\expop$).
We call an operator of radius $\expopr$ symmetric if it uses values of $u$ at 
all grid points within distance $\expopr$ from the target point.
In this section we obtain a lower bound for a symmetric operator of unit radius,
which obviously is a lower bound for symmetric operators of larger radii as well.

\subsection{Pointwise Computations}\label{sec:POINTWISE}
Depending on the separability of the local operator, it has to be computed 
pointwise, or it may allow computation in edgewise fashion.
If an operator is not separable, it requires values of $u$ at all neighbor 
points simultaneously to compute a value of $q$. 
We call such computation \textit{pointwise}.
In this subsection we assume that computation of $q$ on the grid $G$ is 
performed in a pointwise fashion, that is, at any grid point the value of 
$q$ is computed completely before computation of the value of $q$ at another 
point is started.
Separable operators, which can be evaluated edgewise, are considered in the 
next section.

In order to compute the value of $q$ at a grid point $x$,
the values of $u$ at the neighbor points of $x$ must be loaded
into the cache (points $y$ and $x$ are neighbors if 
they are connected by a grid edge). 
If $x$ is a neighbor of $y$ and $u(y)$
has been loaded in cache to compute $q(z)$ but is dropped from
the cache before $q(x)$ is computed, then $u(y)$
must be reloaded, resulting in a replacement load.
Our lower bound on cache loads is based on the following simple observation.

\begin{lemma}\label{lemma:LOWER_BOUND_BASE}
Let values of $u$ at a vertex set $X$ first be used for computation of $q$ in 
a vertex set $U_0$, and next for computation of $q$ in a vertex set $U_1$, $U_0 
\cap U_1 = \emptyset$. 
This computation incurs at least $|X|-\cachec$ replacement loads.
\end{lemma}

\begin{proof}
Since the cache can hold at most $\cachec$ values, at least $|X|-\cachec$ 
values must have been dropped from the cache by the time computations at the 
points of $U_0$ are done.
Since all values at points of $X$ are necessary for computing values at $U_1$,
they have to be reloaded into the cache. 
Hence, computations of values at points of $U_1$ will result in at least 
$|X|-\cachec$ replacement loads.
\end{proof}

\begin{figure}[htb]
\centerline{\epsfig{file=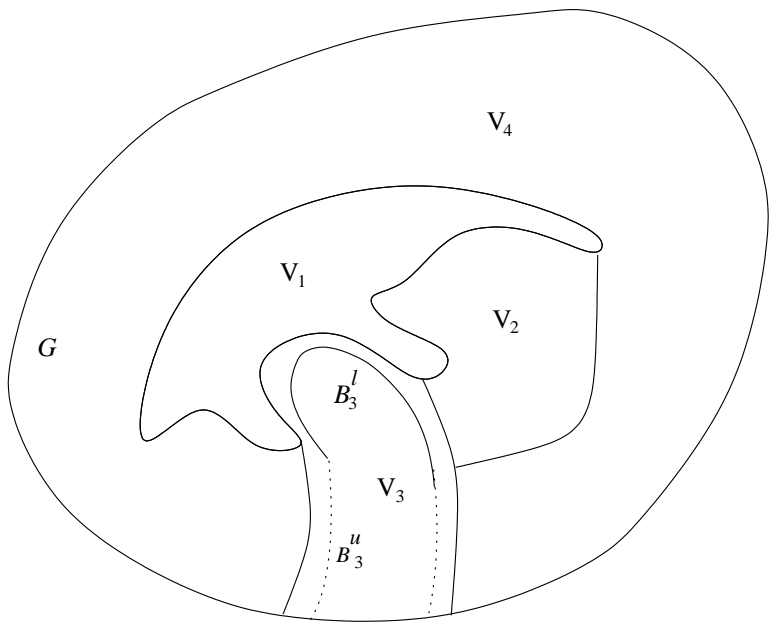,width=4in}}
\caption{\label{fig:ELEPHANT}
Iteration space $V$ is partitioned in a sequence of regions $\vertexsubset_i$.
Reloading of some values of $u$ on the boundary of $\vertexsubset_3$ (heavy and 
dotted lines, corresponding to $|\bdry_{3}^l|$ and $|\bdry_{3}^u|$, respectively) 
results in at least $|\bdry_{3}^l|+|\bdry_{3}^u|-2\cachec$ cache loads.}
\end{figure}

To estimate for a given algorithm the number of elements, $\rho$, of array $u$ 
that must be
replaced, we partition $\vertexsubset$ into a disjoint union of $k$ sets 
$\vertexsubset_i$, with $\vertexsubset = \cup_{i=1}^k \vertexsubset_i$, in such 
a way that $q$
is computed at all points of $\vertexsubset_i$ before it is computed at any 
point
of $\vertexsubset_{i+1}$, see Figure \ref{fig:ELEPHANT}.
Let $\bdry_i^l$ and $\bdry_i^u$ be (possibly overlapping) subsets of 
$\vertexsubset_i$ which have neighbors in $\cup_{j<i}\vertexsubset_j$ and 
$\cup_{j>i}\vertexsubset_j$, respectively. 
The set $\bdry_i=\bdry_i^l \cup \bdry_i^u$ is the boundary of $\vertexsubset_i$.

Replacement loads of values at the points of $\vertexsubset_i$ will be incurred when
such points show up as neighbors in other vertex sets.
We apply Lemma \ref{lemma:LOWER_BOUND_BASE} to the two pairs of sets in which 
boundary points of $\vertexsubset_i$ occur as members and neighbors:
\begin{equation}\nonumber
\cup_{j<i}\vertexsubset_j~\textrm{and}~\vertexsubset_i
\end{equation}
\begin{equation}\nonumber
\vertexsubset_i~\textrm{and}~\cup_{j>i}\vertexsubset_j
\end{equation}
This gives us the following bounds on replacement loads:
\begin{equation}\nonumber
\rho_i^l \geq |\bdry_{i}^l|-\cachec
\end{equation}
and
\begin{equation}\nonumber
\rho_i^u \geq |\bdry_{i}^u|-\cachec \, ,
\end{equation}
respectively.
Hence, in each set of the partition at least $\rho_i$ values have to be reloaded
(see Figure \ref{fig:ELEPHANT}), with 
\begin{equation}\nonumber
\rho_i=\rho_i^l+\rho_i^u \geq |\bdry_i^l|+|\bdry_i^u|-2\cachec \geq 
|\bdry_i|-2\cachec .
\end{equation}

The total number of reloaded values in the course of computing $q$ on the 
entire grid will be at least $\rho=\sum_{i=1}^k\rho_i$, for which we have the 
following bound:
\begin{equation}\nonumber
\rho=\sum_{i=1}^k\rho_i \geq \sum_{i=1}^k |\bdry_i|-2k\cachec.
\end{equation}

Let $v$ be the maximum number of points in a set with $3\cachec$ points 
on its boundary, i.e. let $v$ be the number of points in an isoperimetric set 
with $3\cachec$ boundary points. 
Choosing a partition $\vertexsubset = \cup_{i=1}^k \vertexsubset_i$ in such a way that $|\vertexsubset_i|=v$, we have $|\bdry_i| \geq 3\cachec$, hence
\begin{equation}\label{eq:RHO}
\rho \geq \cachec {\numgp \over v}.
\end{equation}

\noindent
Thus we have the following result.
\begin{theorem}\label{th:LOWER_BOUND_POINTWISE} 
The number of cache misses $\mu$ in the evaluation of a symmetric operator on a 
grid \ugrid is bounded as follows: 
\begin{equation}\nonumber
\mu \geq \frac{1}{\cacheb}(|V|+\rho) \geq \frac{1}{\cacheb}\numgp(1+ \frac{1}{3}\alpha(G))\,
\end{equation}
where  $\alpha(G)$ is the surface-to-volume ratio of isoperimetric subsets 
of $V$ with $3\cachec$ boundary points.
\end{theorem}
\begin{proof}
We sum the number of replacements in (\ref{eq:RHO}) with the number of cold 
loads $\numgp$ and notice that ${\cachec \over v}=1/3\alpha(G)$. 
Then we notice that any cache miss results in a load of $\cacheb$ words into the cache,
and that in the best possible case all $\cacheb$ words contain useful data.
\end{proof}

\subsection{Edgewise Computations}\label{sec:SEPARABLE}

The pointwise calculation model used in the previous subsection is too 
restrictive in many cases. 
Using separability of the operator, the number of cache misses may be reduced.
In this section we present lower bounds for more general computations where 
values of $q$ may be updated multiple times in arbitrary order.
We call these computations edgewise.

An edgewise computation is performed at the vertices of a bipartite 
graph $H=(\vertexset^I,\vertexset^O,\edgeset)$ where, $\vertexset^I$ and 
$\vertexset^O$ are vertex sets, both equal to $\vertexset$, and $(x,y), 
x \in \vertexset^I, y \in \vertexset^O$, is an edge in $H$ iff $x$ and $y$ 
are neighbors in $G$ or $x=y$. 
The values of $u$ are given in the nodes of $\vertexset^I$, and values 
of $q$ have to be computed at the nodes of $\vertexset^O$. 
An atomic step in an edgewise computation is the evaluation of a function 
of two variables corresponding to an edge of $H$. 
If a value at an end point of the edge is currently not loaded, then the 
computation suffers a cache load. 
All edges of the grid must be visited. 
We want to estimate the number of cache loads that each computation of 
$q=\expop u$ must suffer.

The arguments in the proof of the lower bound of Theorem 
\ref{th:LOWER_BOUND_POINTWISE} can be modified by partitioning the edges of 
$H$ into disjoint sets  $\edgsubeset_i$, with 
$\edgeset = \cup_{i=1}^k \edgsubeset_i$, in such a way that processing of any 
edge in $\edgsubeset_i$ precedes processing of any edge in $\edgsubeset_{i+1}$, 
and such that the size of boundary of $\edgsubeset_i$ is at least $3\cachec$. 
Here the boundary of an edge set $\edgsubeset_i$ is the set of vertices 
shared by an edge in $\edgsubeset_i$ and an edge not in $\edgsubeset_i$. 
The surface-to-volume ratio of an edge set is the ratio of its number of 
boundary vertices to the number of its edges. 
We define $\beta (G)$ to be the minimum surface-to-volume ratio of the edge 
sets in $H$ (derived from $G$) having a boundary of size $3\cachec$.
In other words $\beta (G)$ is the number of boundary points divided by the
number of edges in isoperimetric sets in $H$ having $3\cachec$ boundary points.
The following result can be proved using arguments completely analogous to those 
in Theorem \ref{th:LOWER_BOUND_POINTWISE}.

\begin{theorem}\label{th:LOWER_BOUND_SEPARABLE} 
The number of cache misses in a calculation of a separable symmetric 
operator on a grid \ugrid is bounded as follows:
\begin{equation}\nonumber
\mu \geq \frac{1}{\cacheb} |\edgeset| (1+\frac {1}{3}\beta(G)) \,
\end{equation}
where  $\beta(G)$ is the minimum of the surface-to-volume ratio over subsets of $E$ with $3\cachec$ boundary points.
\end{theorem}

\section{Structured Grids}\label{sec:STRUCT_GRIDS}
\subsection{Interference Lattice}\label{sec:INTREFERENCEL}
\textit{Interference lattice.} Let $u$ be a $\grdim$-dimensional array 
defined at the vertices of a structured $\grdim$-dimensional grid of 
size $n_1 \times \cdots\times n_{\grdim}$. Let $L$ be a set in the index 
space of $u$ having the same image in cache as the index $(0,\dots,0)$.
$L$ is a lattice in the sense that there is a generating set of vectors 
$\{{\mathbf b}_i\}$, $i=1,\dots,\grdim$, such that $L$ is the set of grid points 
$\{\sum_{i=1}^\grdim x_i {\mathbf b}_i\, |\, x_i \in {\mathbf Z} \}$.
We call $L$ the \textit{interference lattice} of $u$.
For a direct-mapped cache it can be defined as the set of all vectors 
$(i_1,\dots,i_\grdim)$ that satisfy the Cache Miss Equation \cite{MARTONOSI}:
\begin{equation}\label{eq:CME}
(i_1 + n_1 i_2 + n_1 n_2 i_3 + \cdots + n_1 \cdots n_{\grdim-1} i_\grdim) \bmod \cachec = 0 \, .
\end{equation}

We will use some geometrical properties of lattices. Let $B$ be a convex body of volume $V$, symmetrical about the origin. 
The minimal positive $\lambda_i$ such that $\lambda_i B$ contains $i$ linearly independent vectors of $L$ is called the $i ^{th}$ \textit{successive minimum vector} of a lattice $L$ relative to $B$. 
A theorem by Minkowski, see \cite{CASSELS} (Ch. VIII, Th. V), asserts that
\begin{equation}\label{eq:MINOWSKI}
{2^\grdim \over d! V}\leq {{\prod_{i=1}^d \lambda_i \over \det L}}\leq {2^\grdim \over V}.
\end{equation}
\noindent Note that the ratios of lattice successive minima relative to the unit cube and to the unit ball can be bounded: $1/\grdim \leq \lambda _i ^{cube} / \lambda _i ^{ball} \leq \grdim$. 
We use successive minima relative to the unit ball and the unit cube in the Sections \ref{sec:VORONOI} and \ref{sec:SUCMIN} respectively. 
In both cases we call $f = \lambda _\grdim/ \lambda _1$ the \textit{eccentricity} of the lattice (not to be confused with eccentricity of a reduced basis, defined in \cite{FRUMKIN_WIJNGAART}, Section 4). 
The eccentricities relative to a ball and a cube may differ by a factor of 
$\grdim ^2$ at most. 

In \cite{FRUMKIN_WIJNGAART} we introduced a tiling by parallelepipeds built on 
a reduced basis of the interference lattice which decreases the number of the 
cache misses to a level close to the theoretical lower bound that we also derived. 
Measurement shows that this tiling incurs significantly fewer cache misses than a compiler-optimized code with canonical loop ordering. 
However, it has a high computational cost, since it depends on a significant 
number of integer parameters, and its implementation scans a substantial 
number of the grid points to select those suitable for cache conflict-free 
computations. 
This prompts us to consider alternative tilings, specifically with Voronoi 
cells and with successive minima parallelepipeds. 
We show that these tilings have good surface-to-volume ratios if the lattice 
eccentricity is small.

\subsection{Voronoi Tiling}\label{sec:VORONOI}
\textit{A Voronoi tiling} is a tiling of the grid by completed cells $C$ (Voronoi tile) of the Voronoi diagram of the interference lattice. 
For each lattice point $x$ a Voronoi cell is the set of points which are closer to $x$ than to any other lattice point. 
All integer points inside each Voronoi cell are mapped into the cache without conflicts. 
Voronoi cells may not form a tiling since some integer points can be located on a cell boundary. 
There are many qualitatively equivalent ways to complete the cells to form a tiling. 
One way is to choose a basis in the space of the lattice and assign an 
integer point to the cell whose center is lexicographically closest to the point.
   
In order to estimate the surface-to-volume ratio of a Voronoi cell $C$ we note that the completed Voronoi cells form a tiling of space. 
Hence, the volume of $C$ equals the determinant of the lattice (defined 
by (\ref{eq:CME})), which is equal to $S$, see \cite{FRUMKIN_WIJNGAART}. 
On the other hand, let $C_o$ be a Voronoi cell centered at  $o$.
Each vertex $v$ of $C_o$ is equidistant from $d$ lattice points. 
Let $r$ be that distance, which will in general be different for different $v$. 
According to the definition of the Voronoi cell, the ball of radius $r$, 
centered at $v$, contains no other lattice points. 
Hence $r \leq R$, where $R$ is the radius of a \textit{maximal ball} of the 
lattice (a lattice-points-free ball of maximal radius).
Hence, $C_o$ is contained in a ball of radius $R$, centered at $o$.
Thus, the surface area of $C_o$ is bounded by the surface of a 
ball of radius $R$, which equals $\grdim V_d\ R ^{\grdim-1}$, where 
$V_\grdim= {\pi ^{\grdim/2} \over \Gamma(1+\grdim/2)}$ is the volume of the unit 
$\grdim$-dimensional ball (see \cite{CASSELS} Ch. IX.7).

\begin{figure}[htb]
\centerline{\epsfig{file=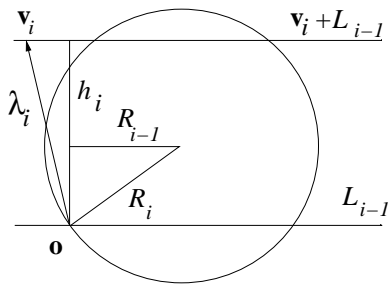,width=2in}}
\caption{\label{fig:INDUCTIONFIG}
Let $L_i$ be a lattice generated by the first $i$ minima vectors of $L$. 
An intersection of a maximal ball of $L_i$ with the subspace of $L_{i-1}$ 
is contained in a maximal ball of $L_{i-1}$. 
Hence $R_i ^2 \leq (h_i/2)^2 +R_{i-1}^2$ where $h_i$ is the distance between 
subspaces $L_{i-1}$ and ${\mathbf v}_i+L_{i-1}$.%
}
\end{figure}

We estimate the radius of the maximal ball $R$ by induction on the dimension of 
a sublattice. 
Let $R_i$ be the radius of the maximal ball inscribed into the lattice $L_i$ 
generated by the first $i$ minima vectors ${\mathbf v}_i$ of $L$ relative to the
$i$-dimensional ball.
Then we have (see Figure \ref{fig:INDUCTIONFIG}):
\begin{equation}\nonumber
R_i ^2 \leq (h_i/2)^2 +R_{i-1}^2 \leq (\lambda_i/2)^2 +R_{i-1}^2 \, ,
\end{equation}
where $h_i$ is the distance between hyperplanes spanned by $L_{i-1}$ and
$L_{i-1} + {\mathbf v}_i$, respectively, and $\lambda_i$ is the $i^{th}$ successive
minimum.
Induction on $i$ gives the following result.

\begin{lemma}
The following relation holds for the radius of a lattice-points-free ball 
in a $\grdim$-dimensional lattice $L$:
\begin{equation}\label{eq:RAD_MAX_BALL} 
R ^2 \leq 1/4\sum_{i=1}^\grdim \lambda_i ^2 . 
\end{equation}
where $\lambda_1 \leq \dots \leq \lambda_\grdim$  are successive minima of $L$.
\end{lemma}

Since $C$ is contained in a ball of radius $R$ and both $C$ and ball are 
convex bodies, the surface area $A$ of $C$ does not exceed the surface area 
of the ball and we have an estimation
\begin{equation}\nonumber
A(C)\leq 
\grdim V_\grdim R^{\grdim-1} \leq 
\grdim(\sqrt \grdim/2)^{\grdim-1} V_\grdim {\lambda _d}^{\grdim-1} \leq
\grdim^{{\grdim+1}\over 2} V_\grdim ^{1/\grdim} f^{(\grdim-1)^2 /\grdim} S^{(\grdim-1)/\grdim},
\end{equation}
where we used the estimation $\lambda _\grdim ^\grdim \leq {2^\grdim \over V_\grdim} f^{\grdim-1}S $ derived from \eqref{eq:MINOWSKI}, and the bound $R \leq {\sqrt \grdim \over 2} \lambda _\grdim$ which follows from \eqref{eq:RAD_MAX_BALL}.
This implies the following result.
\begin{theorem}\label{th:SURFVOL_VORONOI}
The surface-to-volume ratio of a lattice Voronoi cell $C$ can be estimated as:
\begin{equation}\nonumber
{A(C)\over V(C)} \leq c_\grdim f^{(\grdim-1)^2/\grdim} \cachec ^{-1/\grdim}
\end{equation}
where $c_\grdim$ is a constant depending on $\grdim$ only, and $f$ is the
lattice eccentricity.
\end{theorem}

\subsection{Successive Minima Tiling}\label{sec:SUCMIN}

The Voronoi cell tiling has cache-conflict-free tiles of maximum possible 
volume $\cachec$, and of small surface-to-volume ratio. 
However, the tiles may have many faces, and it may be computationally expensive 
to traverse the grid points inside a tile. 
It is more desirable to use rectilinear tiles. 
A \textit{successive minima tiling} is a tiling by Cartesian blocks 
constructed using successive minima lattice vectors of the unit cube. 
Such a block $Q$ can be described by the system

\begin{equation}\label{eq:PARALLELEPIPED}
|x_i| \leq b_i,~i=1,\dots ,\grdim ,
\end{equation}
where $\lambda_1 \leq b_i \leq \lambda_\grdim$. 

The block $Q$ can be constructed by the following ``inflating'' process. 
Take an initial cube of the form \eqref{eq:PARALLELEPIPED}, with 
$b_i=1,~i=1,\dots ,\grdim$, and increment all $b_i$ equally until the face 
$x_{i_1}=b_{i_1}$ contains a lattice point. 
Continue to increment values of all $b_j$ for which the face $x_j=b_j$ has 
no lattice points. 
At the end we obtain a block of the form \eqref{eq:PARALLELEPIPED}, containing a lattice point on each of its faces and containing no lattice points inside except $o$. 
If each successive minimum vector has end points in distinct faces of the 
block, then  $b_i=\lambda_i$ (after an appropriate reordering of the coordinates). 
On the other hand, it is not difficult to construct a three-dimensional 
lattice such that the block satisfies $b_1=\lambda_1,~b_2=b_3=\lambda_2<\lambda_3$, 
in which case the volume of the block would be strictly less than 
$\lambda_1 \cdots \lambda_\grdim$.

Any translation of the block $Q ^\prime ={1\over 2}Q$ obviously contains at most one lattice point and can be used for conflict free tiling. 
This block has a low surface-to-volume ratio if the lattice has bounded eccentricity, which can be seen from the following inequalities: $A(Q^\prime) \leq 2\grdim \lambda _\grdim ^{\grdim-1}$ and $V(Q^\prime) \geq \lambda_1^\grdim$. 
Since $\lambda_1 = \frac{\lambda_\grdim}{f} 
\geq \frac{2}{f}({S\over {\grdim!V_\grdim}})^{1/\grdim}$, as follows from 
\eqref{eq:MINOWSKI}, we obtain the following result.
\begin{theorem}\label{th:SURFVOL_SUCMIN}
The surface-to-volume ratio of the block $Q^\prime$ obtained in a successive
minima procedure can be estimated as follows:
\begin{equation}\nonumber
{A(Q^\prime) \over V(Q^\prime)} \leq 
2\grdim f ^{\grdim-1} /{\lambda _1} \leq 
\grdim(\grdim! V_\grdim)^{1/\grdim} f ^\grdim \cachec ^{-1/\grdim}  \, ,
\end{equation}
where $V_\grdim$ is a constant depending on $\grdim$ only, and $f$ is the 
lattice eccentricity.
\end{theorem}

\begin{figure}[htb]
\centerline{\epsfig{file=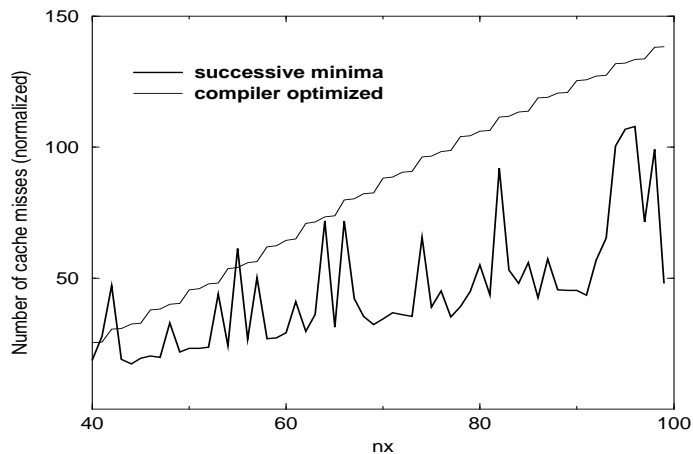,height=2.4in,width=3.6in}}
\caption{\label{fig:CACHE_MISS_COUNT} 
Comparison of measured cache misses for the second order stencil operator as a 
function of the first grid dimension ($40 \leq nx \leq 99,~ny=97,~nz=99$). 
The top curve shows the number of cache misses for the compiler optimized 
canonically ordered nest. 
The bottom curve is obtained for tilings with successive-minima parallelepipeds.
}
\end{figure}
As a representative example, the number of measured primary cache misses for 
tilings of three-dimensional grids of sizes $40 \leq nx \leq 99,~ny=97,~nz=99$ 
with successive minima parallelepipeds is shown in Figure \ref{fig:CACHE_MISS_COUNT}. 
The computation evaluates a linear operator on the standard 13-point star stencil.
Experiments were performed on an SGI Origin 2000 machine with a MIPS R10000 processor.

It is easy to show that the lattice eccentricities $f$ can be bounded by the
eccentricity of a reduced basis as defined in \cite{FRUMKIN_WIJNGAART}, times a
constant depending on $\grdim$ only.
Hence, it follows from Theorems \ref{th:SURFVOL_VORONOI} and \ref{th:SURFVOL_SUCMIN}
that the Voronoi and successive minima tilings have the same asymptotic 
surface-to-volume ratio---and therefore the same cache efficiency---as the tiling
with fundamental parallelepipeds derived in \cite{FRUMKIN_WIJNGAART}.
\section{Unstructured Grids}\label{sec:UNSTRUCT_GRIDS}
\subsection{Lipton-Tarjan Covering}\label{sec:LIPTON}
In this section we present a covering of a planar bounded-degree grid with 
vertex sets of size at most $S$ that have an average surface-to-volume ratio equal 
to ${\cal O}(1/\sqrt \cachec)$. 
The covering can be used for computing a first-order operator on
a $|V|$-vertex planar grid with ${\cal O}(|V|/ \sqrt{\cachec})$ replacement loads. 
The covering is based on the Lipton-Tarjan separator theorem, which asserts that 
any planar graph of $|V|$ vertices has a vertex separator of size ${\cal O}(\sqrt{|V|})$. 
The separator can be constructed in ${\cal O}(|V|)$ time \cite{LIPTON}.

We consider bounded-degree unstructured grids, that is, grids having fixed 
maximum vertex degree $d$, independent of the grid size. 
(Calculation on unbounded-vertex-degree grids may cause approximation problems 
and numerical instability. 
Consequently, most numerical methods employ bounded-degree grids.) 
For bounded degree grids a node cut of size ${\cal O}(|V|)$ has a corresponding edge 
cut of size ${\cal O}(|V|)$, and vice versa. 
Hence, the Lipton-Tarjan separator theorem 
(see \cite{LIPTON}, section 2, Corollary 2) can be reformulated as follows: 
By removing ${\cal O}(\sqrt{|V|})$ edges from a $|V|$-vertex bounded-degree planar 
graph it can be separated into connected disjoint subgraphs, each having at most 
$2|V|/3$ vertices.  

We construct the covering by applying the Lipton-Tarjan theorem recursively. 
First, we choose an edge cut $C$ of the original graph $G=(V,E)$, such that 
$|C| \leq c_0 \sqrt |V|$, where $c_0$ is independent of $|V|$. 
According to the Lipton-Tarjan separator theorem, the cut can be chosen in such 
a way that it will split the graph into connected components
$G_i=(V_i,E_i),~~i=1, \dots ,k,~|V_i| \leq 2|V|/3$. 
Adding an extra step in this partition, we can assume that $|V_i| \leq |V|/2$ 
while $|C| \leq c_1 \sqrt |V|$ for a bigger constant $c_1$. 
Then, we recursively bisect each connected component $G_i=(V_i,E_i)$ while $|V_i| > \cachec$.
We call this covering a \textit{Lipton-Tarjan covering}.

This partition process can be represented by a cut tree $T$ whose nodes 
are partitioned connected components of the grid. 
The tree will be used to estimate the total number of removed edges to 
construct a Lipton-Tarjan covering. 
Let a connected component $C$, represented by node $t$ of $T$, be split into 
components $C_1,\dots,C_m$ by a Lipton-Tarjan cut. 
We draw an edge between $t$ and each of its child nodes representing $C_1,\dots,C_m$.
However, we do not include in $T$ the connected components of size smaller than $S$.
Hence, the size of each leaf (i.e.\ a node having no children) exceeds $\cachec$.
Connected components smaller than $\cachec$ do not contain any cut edges and 
they are not needed for the estimate. 
To each node $t$ of $T$ we assign \textit{size} $s(t)$, which equals the number of 
vertices in the set represented by $t$, and \textit{weight} $w(t)=\sqrt {s(t)}$. 

\begin{lemma}\label{lm:CUTLEMMA}
The total number of edges in all cuts of a Lipton-Tarjan covering is 
${\cal O}(|V| / \sqrt S)$.
\end{lemma}
\begin{proof} 
The total number of edges in all cuts made to construct a Lipton-Tarjan 
covering can be bounded by ${\cal O}(\sigma (T))$, where
\begin{equation} \label{eq:weight_sum}
\sigma (T)=\sum_{t~\mbox{\textrm{\scriptsize node of}}~T} w(t)
\end{equation}

\noindent
We use two properties of the weights. First,
\begin{equation} \label{eq:leaf_sum}
\sum_{l~\mbox{\textrm{\scriptsize leaf of}}~T} w(l) \leq |V| / \sqrt \cachec,
\end{equation}
\noindent 
since the maximum of the $\sum \sqrt {s(l)} $, subject to 
$\sum s(l)=|V| ~\textrm{and}~ s(l) \geq \cachec+1$, is attained for 
$s(l)=\cachec+1$ for all $l$. 
Second, for each node $t$ of $T$ we have
\begin{equation} \label{eq:geom_eneq}
w(t)<1/ \sqrt 2 \sum_{\tau~\mbox{\textrm{\scriptsize child of}}~t} w(\tau),
\end{equation}
which follows from Proposition \ref{th:convex_roots_d} with $\grdim=2$, 
see Appendix.

Now $\sigma(T)$ can be estimated in two steps. 
First, we replace weights in each nonleaf $t$ by the right hand side of 
\eqref{eq:geom_eneq}, going bottom up from the leaves to the root of $T$. 
This operation will not decrease the total weight. 
Second, we carry summation of new weights across nodes of $T$ by noticing that 
each leaf $l$ deposits into $\sigma(T)$ a weight of at most
\begin{equation}\nonumber
w(l)(1 + {1 \over {\sqrt 2}} +{1\over {\sqrt {2^2}}} + \dots ) \leq w(l)(2+\sqrt 2).
\end{equation}
\noindent Hence, it follows from \eqref{eq:leaf_sum} that
\begin{equation} \label{eq:final_estim}
\sigma (T) \leq(2+ \sqrt 2)|V|/\sqrt \cachec,
\end{equation}
so the total number of edges in all cuts is ${\cal O}(|V| / \sqrt S)$.
\end{proof}

Now we prove the main result of this section. 
From a lower bound on the number of replacement loads for computations of 
star stencil operators on structured two-dimensional grids, see Section \ref{sec:LOWER}, 
cf.\ \cite{FRUMKIN_WIJNGAART}, it follows that the order of this bound can not be improved.
\begin{theorem}
Any  first-order operator $K$ on a bounded-degree planar grid $G=(V,E)$ 
can be computed with ${\cal O}(|V|/\sqrt \cachec)$ replacement loads.
\end{theorem}
\begin{proof}  
First, we reorder vertices of the grid in such a way that vertices of each set of 
the Lipton-Tarjan covering will occupy contiguous memory locations. 
Then we compute $K$ on each set, one at a time. 
In this computation replacement loads can happen only at the vertices of the cuts 
of the grid. 
According to Lemma \ref{lm:CUTLEMMA} the number of such vertices does not exceed 
${\cal O}(|V|/\sqrt \cachec)$.
\end{proof}

\subsection{Covering of \Uniformg\ Grids}\label{sec:UNIFORM}
In the previous sections we showed that structured and planar grids can be 
covered by sets having low surface-to-volume ratios.
That implies that local operators of fixed order on such grids can be 
evaluated with high efficiency of cache utilization.
In this section we introduce a class of multi-dimensional unstructured grids, which 
we call \uniformg\ grids.
We show that cache efficiency of \uniformg\ grids is the same as that 
of structured grids.

\begin{definition}
We call a grid a $\grdim$-dimensional \textit{\uniformg} grid if it can be mapped to 
a grid in $\grdim$-dimensional Euclidean space that has the following two properties:

\begin{itemize}
\item The ratio of the length $L$ of the longest edge of the grid 
to the length $l$ of the shortest edge of the grid is limited by a constant 
independent of the number of grid points:
\begin{equation}\label{eq:EDGE_RATIO}
L \leq \const_0 l \, .
\end{equation}
\item There are no grid points at a distance shorter than $l$.
\end{itemize}
\end{definition}

A vertex of a \uniformg\ grid can be adjacent only to grid points contained in 
a ball of radius $L$ centered at this point. 
On the other hand, a ball of radius $l$ around any vertex of a \uniformg\ 
grid is free of other grid points. 
Hence, the vertex degree of a $\grdim$-dimensional \uniformg\ grid is bounded by 
a constant depending on $\grdim$ only.
Another property of \uniformg\ grids is that any subgrid is \uniformg\ as well.
Any grid with $|\vertexset|$ vertices is a $(|\vertexset|-1)$-dimensional 
\uniformg\ grid, since it can be mapped onto the vertices of a unit 
$(|\vertexset|-1)$-dimensional simplex.
In this paper, however, we consider $\grdim$ to be fixed, and $|\vertexset|$ to be 
a parameter.

\Uniformg\ grids are common in many computations, such as those involving 
particles distributed in a finite domain and interacting via a short range potential. 
Unstructured discretization grids for solving partial differential equations
are often \uniformg\ as well.
While there is no obvious way to verify that an arbitrary grid is a $\grdim$-dimensional 
\uniformg\ grid, 
starriness of a grid embedded into Euclidean space can easily be verified. 
One simple way to construct a \uniformg\ grid is to delete some vertices and their
incident edges from a structured grid. 

We will show that a $\grdim$-dimensional \uniformg\ grid \ugrid can be covered 
by sets of size $\cachec$ having at most $\const |\vertexset|S^\dmin$ boundary 
points in total, see Theorem \ref{th:covering_d}. 
This covering, as in the case of planar grids, is based on a cutting theorem, 
specifically, the \hcutth\ (Theorem \ref{th:Hyperplane_Cut_Theorem}), 
asserting that there is a hyperplane bisecting $\vertexset$ into almost 
equal sized parts while cutting at most $\const |\vertexset|^\dmood$ edges 
of the grid. From this we deduce the main result of the section.

\begin{theorem}
Any first-order operator $\expop$ on a \uniformg\ $\grdim$-dimensional grid \ugrid can be computed with 
${\cal O}(|\vertexsubset|\cachec^{-\frac{1}{\grdim}})$ replacement loads.
\end{theorem}

\begin{proof}
First, we reorder vertices of the grid in such a way that the vertices of 
each set of the covering occupy contiguous memory locations. 
Then we compute $\expop$ within each set separately. 
In this computation replacement loads can happen only at the vertices of the cuts of the grid. 
According to Theorem \ref{th:covering_d} the number of such vertices does not 
exceed ${\cal O}(|\vertexsubset| \cachec^{-\frac{1}{\grdim}})$.
\end{proof}

From a lower bound on the number of replacement loads for computations of  operators 
on structured $\grdim$-dimensional grids, see Section \ref{sec:LOWER}, it follows 
that this bound can not be improved. 

It is not difficult to see that any continuous convex body has a small 
bisector (for example, a hyperplane orthogonal to a diameter of the body). 
It is not surprising, then, that if a grid represents the body well, it has a 
small bisector size as well.
We will construct a set containing $3\grdim$ vectors, and show that the 
bisecting hyperplane can be chosen to be normal to one of these vectors. 
This constitutes an algorithm of complexity ${\cal{O}}(|\vertexset|^2)$ for finding 
such a bisector.

The following is a multi-dimensional analog of the Lipton-Tarjan cut theorem. 

\begin{theorem}\label{th:Hyperplane_Cut_Theorem}
(\hcutth) For any \uniformg\ $\grdim$-dimensional grid \ugrid  
there is a hyperplane cut of size ${\cal O}(\numgp^{\frac{\grdim-1}{\grdim}})$ 
separating vertices of the grid into two sets of size at least $\numgp/3$.
\end{theorem}

\begin{proof} 
Let us consider vectors of unit length ${\mathbf u}_i \in 
{\cal R}^\grdim,~|{\mathbf u}_i|=1,~i=1,\dots ,k$ such that all vertices have different
orthogonal projections onto each ${\mathbf u}_i$.
Then we choose slabs $S_i$ bounded by hyperplanes orthogonal to ${\mathbf u}_i:~S_i=
\{{\mathbf x}\in
{\cal R}^\grdim | \lambda _i\leq {\mathbf u}_i {\mathbf x} \leq \mu _i\}$, 
trisecting $V$, see Figure
\ref{fig:FINITE_RADON}, that is
\begin{equation}\label{eq:TRISECTION}
|\{{\mathbf v} \in \vertexset| {\mathbf u}_i {\mathbf v} \leq \lambda _i\}| 
\geq \numgp/3 ~\textrm{while}~
|\{{\mathbf v} \in \vertexset| {\mathbf u}_i {\mathbf v} < \lambda _i\}| < \numgp/3,
\end{equation}
\begin{equation}\nonumber
|\{{\mathbf v} \in \vertexset| {\mathbf u}_i {\mathbf v} \geq 
\mu _i\}| \geq \numgp/3 ~\textrm{while}~ |\{{\mathbf v}
\in \vertexset| {\mathbf u}_i {\mathbf v} > \mu _i\}| < \numgp/3.
\end{equation}
meaning that each slab contains at least $\numgp/3$ points, while at most
$\numgp/3$ grid points can be contained strictly inside a slab.
We will show that ${\mathbf u}_1,\dots ,{\mathbf u}_k$ can be chosen in such a 
way that at least one slab is wide, that is 
$ h_i=\mu _i - \lambda _i \geq {\cal O}(\numgp^{1/\grdim}l)$.
If $|\vertexset|$ is big enough, then it follows from (\ref{eq:EDGE_RATIO}) that 
no grid edge can pierce this slab, hence there exists a hyperplane
$H_i=\{{\mathbf x}:{\mathbf u}_i {\mathbf x} = \eta _i\}, ~ 
\lambda _i< \eta _i< \mu _i$ that intersects at most 
${\cal O}(\numgp ^{\frac{\grdim-1}{\grdim}})$ grid edges.
This hyperplane separates grid points into sets containing at least $\numgp/3$
points.
\begin{figure}[htb]
\centerline{\epsfig{file=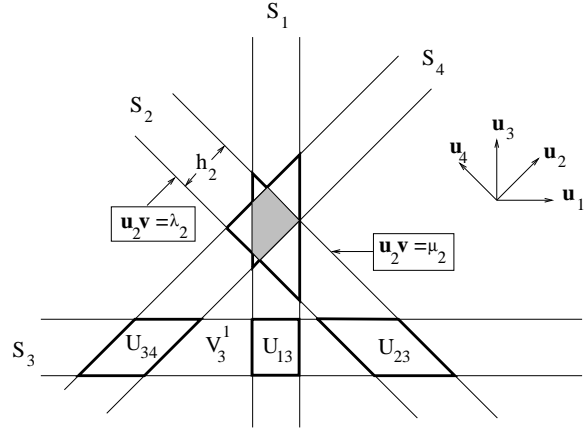,width=3.0in}}
\caption{\label{fig:FINITE_RADON} Finding a wide bisecting slab. 
$\vertexsubset ^2$ is the set of points contained in the thick-lined polygons,
minus those contained in $\vertexsubset ^3$, the shaded polygon.}
\end{figure}

Let $\vertexsubset^i$ be the set of grid points contained in exactly $i$ slabs, 
and let $\vertexsubset_j^i$ be the subset of $\vertexsubset^i$ contained in $S_j$. 
Since $\vertexsubset^i~,i=1,\dots,k$, are disjoint sets and 
$\cup_{i\leq k}\vertexsubset^i \subseteq \vertexset$, we have 
\begin{equation}\label{eq:NUM_GR_POINTS_INEQ}
\sum |\vertexsubset ^i| \leq \numgp.
\end{equation}
Since each slab contains at least $|\vertexset|/3$ grid points, it follows that 
\begin{equation}\nonumber
\sum_j |\vertexsubset _j^i| \geq \numgp/3.
\end{equation}
If we sum all points in all slabs, then each point in $\vertexsubset^i$ 
will be counted exactly $i$ times, hence
\begin{equation}\nonumber
\sum _{1\leq i \leq k}i|\vertexsubset ^i| \geq \frac{k}{3}\numgp ,
\end{equation}
and
\begin{equation}\label{eq:LB_ON_GP_IN_INTERSCT}
\sum_{i \geq d} i|\vertexsubset ^i| \geq \frac{k}{3}\numgp-\sum_{i < 
\grdim} i|\vertexsubset ^i| \geq (\frac{k}{3}-\grdim+1)\numgp \, ,
\end{equation}
where the last inequality follows from (\ref{eq:NUM_GR_POINTS_INEQ}).
Let $\ppp _I,~I=\{i_1,\dots ,i_\grdim\},1 \leq i_1< \dots <i_\grdim \leq k$,
be a parallelepiped formed by the intersection of $d$ different slabs. 
Obviously, $\cup _{i \geq \grdim} V^i \subseteq \cup _I \ppp _I$. 
Let $\grpppp _I$ be the number of grid points in $\ppp _I$, then
\begin{equation}\nonumber
\sum _I \grpppp _I \geq \sum _{i \geq \grdim} {{i}\choose {\grdim}} 
|\vertexsubset^i| \geq 1/d \sum _{i \geq \grdim} i |\vertexsubset^i| \, .
\end{equation}\nonumber
If we now choose $k=3\grdim$, it follows from (\ref{eq:LB_ON_GP_IN_INTERSCT}) that
\begin{equation}\label{eq:LB_ON_GP_IN_INTERSCT1}
\sum _I \grpppp _I \geq \frac{1}{\grdim}\numgp. 
\end{equation}
According to the second property of \uniformg\ grids, the balls
$B_l$ of radius $l$, centered at the grid points, do not intersect, 
so 
\begin{equation}\label{eq:LB_ON_PP_VOLUME}
\sum _I \grpppp _I \leq (\const_3 l^\grdim)^{-1} \sum _I \textrm{vol}(\ppp _I + B_l), 
\end{equation}
where $\const_3 l^\grdim = \textrm{vol}(B_l)$.

Now we exercise our freedom to select the vectors ${\mathbf u}_i$. 
We choose them to be the rows (normalized to unit length) of a 
$3\grdim \times \grdim$ generalized Vandermonde matrix 
\begin{equation}\nonumber
\vandermond = | (i-\alpha)^j|,~i=1,\dots ,3d,~j=0,\dots ,\grdim -1,~0 < \alpha < 1
\end{equation}
and choose $\alpha$ in such a way that ${\mathbf u}_i({\mathbf v}_p-{\mathbf v}_q) \neq 0$, 
where ${\mathbf v}_p,{\mathbf v}_q$ are different grid points.
These gives us at most $3\grdim|\vertexset|^2$ nontrivial constraints.
We choose $\alpha$ such that inequality holds for all of them.
Then each parallelepiped $\ppp _I$ may be described by a system of inequalities 
\begin{equation}\nonumber
-h_I \leq D_I \vandermond _I({\mathbf x}-{\mathbf x}_I) \leq h_I \, ,
\end{equation}
where $\vandermond _I$ is a square matrix, consisting of rows $(i_1,\dots ,i_\grdim)$ 
of $W$, and $h_I=(h_{i_1},\dots,h_{i_\grdim})^{t}$, $h_{i_s}$ is the halfwidth 
of $\ppp _I$ in the direction of the vector ${\mathbf u}_{i_s}$, ${\mathbf x}_I$ is the 
center point of $\ppp _I$ and $D_I$ is the normalizing matrix. 
Hence, $\ppp _I + B_l$ is contained in a ball of radius $R_I$, with
\begin{equation}\nonumber
R_I = |D_I^{-1}||\vandermond _I^{-1}| |h_I + l| \leq (3\grdim)^\grdim\sqrt \grdim 
\prod_{i,j\in I,i < j} (i - j) (h+l) < 3^{2\grdim} \grdim^{2\grdim+1/2}(h+l) \, ,
\end{equation}
where $h=\max_{1\leq i\leq 3\grdim}\{h_i\}$ and all matrix norms 
are those induced by the Euclidean distance. 
If $h \leq l$, then each  $\ppp _I$ contains at most one grid point, which 
contradicts (\ref{eq:NUM_GR_POINTS_INEQ}) for $|\vertexset|$ large enough, hence  
$h>l$, and we have
\begin{equation}\nonumber
\textrm{vol}(\ppp _I + B_l) \leq (h+l)^{\grdim}\textrm{vol}(\ppp _I) \leq c_4 h^{\grdim}.
\end{equation}
From this inequality, together with (\ref{eq:LB_ON_GP_IN_INTERSCT1}) and 
(\ref{eq:LB_ON_PP_VOLUME}), it follows that
\begin{equation}\nonumber
\const_5 \numgp \leq \left( \frac{h}{l} \right)^{\grdim}.
\end{equation} 
This results in the desired lower bound for the width of a slab containing 
at least $\numgp /3$ grid points
\begin{equation}\nonumber
h \geq \const_6 \numgp^{1/\grdim}l.
\end{equation} 

Now we show that there is a hyperplane parallel to the boundaries of the slab 
which intersects at most $\const_6 \numgp ^{\frac{\grdim-1}{\grdim}}$ edges 
of the grid. 
We divide the slab into $\lfloor \frac{h}{L} \rfloor \geq \const_6 \frac{l}{L} 
\numgp^{1/\grdim} \geq \const_7 \numgp^{1/\grdim}$ parallel slices of width $L$. 
The number of these slices is bigger than 1 if $\numgp$ is sufficiently large.
Since the total number of grid points in the slab does not exceed $\numgp$, at 
least one slice contains fewer than $\const_8 \numgp ^{\frac{\grdim-1}{\grdim}}$ 
grid points. 
If a grid edge intersects bisector $H$ of the slice, then at least one end of 
the edge is inside the slice (since the slice has thickness $L$ and the 
length of no edge exceeds $L$). 
Hence, the total number of the edges intersecting $H$ does not exceed  
$\const_8 \numgp ^{\frac{\grdim-1}{\grdim}} \delta$, where $\delta$ is the 
degree of the grid, which is bounded for \uniformg\ grids. 
Since $H$ is inside the slab, it separates the grid points into sets containing 
at least $\numgp /3$ points each.
\end{proof}

Now we can formulate our covering result, which implies that computations 
of local operators of fixed order on \uniformg\ grids can be performed with the 
same cache efficiency as those on structured grids.

\begin{theorem}\label{th:covering_d}
Nodes of  $\grdim$-dimensional \uniformg\ grid \ugrid  
can be covered by sets of size not exceeding $\cachec$ and with total boundary 
${\cal O}(|\vertexset|\cachec^{-\frac{1}{\grdim}})$.
\end{theorem}

\begin{proof}
The proof closely follows that of the main result of Section \ref{sec:LIPTON}. 
As was stated in the beginning of the section, any subgrid of a 
\uniformg\ grid is \uniformg\ of the same dimension. 
Hence, we can construct the covering by applying the \hcutth\ recursively. 
First, we choose any bisector cutting at most $c_0 |\vertexset|^\dmood$ 
edges of the grid $G$, where $\const_0 $ is independent of $G$. 
According to the \hcutth, the bisector can be chosen in such a way that it 
splits the grid into connected components
$G_i=(\vertexsubset _i,E_i),~~i=1, \dots ,k,~|\vertexsubset _i| \leq 2|\vertexsubset| /3$. 
Adding an extra step in this partition, we can assume that 
$|\vertexsubset _i| \leq |\vertexsubset| /2$, while the number of edges 
cut by the bisector does not exceed $\const_1 |\vertexsubset|^\dmood $ for 
a bigger constant $\const_1$. 
We recursively bisect each connected component 
$G_i=(\vertexsubset _i,E_i)$ while $|\vertexsubset _i| > \cachec$.

This partition process can be represented by a cut tree $\cuttree$ 
whose nodes are partitioned connected components of the grid. 
Let a connected component $C$, represented by node $t$ of $T$, be split 
into components $C_1,\dots,C_m$ by a hyperplane cut. 
We draw an edge between $t$ and each of its child nodes representing $C_1,\dots,C_m$.
However, we do not include in $\cuttree$ connected components of size smaller 
than $\cachec$. 
To each node $t$ of $\cuttree$ we assign size $s(t)$, which equals the 
number of vertices in the set represented by $t$, and weight $w(t)=s(t)^\dmood$. 
From the definition of the cut tree it follows that the size of each leaf exceeds $\cachec$. 
The total number of edges in all cuts can be bounded by ${\cal O}(\sigma (\cuttree))$, where
\begin{equation} \label{eq:weight_sum_d}
\sigma (\cuttree)=\sum_{t~\mbox{\texttt{\scriptsize node of}}~\cuttree} w(t)
\end{equation}

\noindent
We use two properties of the weights. First,
\begin{equation} \label{eq:leaf_sum_d}
\sum_{l~\mbox{\textrm{\scriptsize leaf of}}~\cuttree} w(l) 
\leq |\vertexsubset|  \cachec^\dmin \, ,
\end{equation}
\noindent 
since the maximum of $\sum {s(l)^\dmood} $ for sizes of the nodes, 
subject to  $\sum s(l)=|\vertexsubset|$ and $s(l) \geq \cachec+1$, is attained 
at $s(l)=\cachec+1$ for all $l$. 
Second, as follows from Proposition \ref{th:convex_roots_d} (see Appendix), for 
any node $t$ of $T$ we have
\begin{equation} \label{eq:geom_eneq_d}
w(t)< \const \sum_{\tau~\mbox{\textrm{\scriptsize child of}}~t} w(\tau)
\end{equation}
for some $\const < 1$ independent of the grid size. 

Now $\sigma(\cuttree)$ can be estimated in two steps. 
First, we replace weights in each nonleaf $t$ by the right hand side of 
\eqref{eq:geom_eneq_d}, going bottom up from the leaves to the root of $\cuttree$. 
This operation will not decrease the total weight. 
Second, we carry the summation of the new weights across nodes of $\cuttree$ by 
noticing that each leaf $l$ deposits into the total sum $\sigma (\cuttree)$ 
a weight of at most
\begin{equation}\nonumber
w(l)(1+\const+\const^2+ \dots )= w(l)\const_5.
\end{equation}
\noindent 
Hence, it follows from \eqref{eq:leaf_sum_d} that
\begin{equation} \label{eq:final_estim_d}
\sigma (\cuttree) \leq \const_5|\vertexsubset|\cachec^\dmin\, ,
\end{equation}
meaning that the total number of edges in all cuts is 
${\cal O}(|\vertexsubset|  \cachec^\dmin)$.
\end{proof}

\subsection{Cache Unfriendly three-dimensional Grid}\label{sec:CACHE_UNFRIENDLY}
In the previous section we provided a characterization of cache-friendly 
higher-dimensional unstructured grids, called \uniformg\ grids.
These are not a trivial extension of planar, bounded degree grids to
higher dimensions.
To demonstrate this, we present a three-dimensional grid
of bounded degree that is intrinsically cache-unfriendly.
More formally, we construct a three-dimensional grid of $N$ vertices which has 
a subgrid $G$ of size $cN$ that does not contain small subsets with a small 
surface-to-volume ratio. 
Using this property, following the arguments of Section \ref{sec:LOWER}, it 
can be shown that for any computation of a symmetric operator defined on this
grid, $\Omega (N/\log S)$ replacement misses must occur.
 
\begin{figure}[htb]
\centerline{\epsfig{file=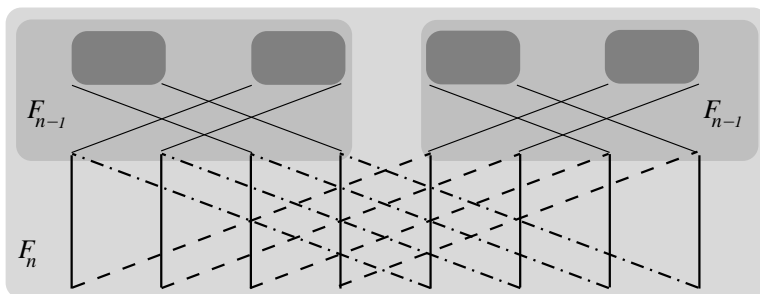,width=4in}}
%\centerline{\epsfig{file=butterfly.eps,width=4in}}
\caption{\label{fig:FFT_NETWORK_FIG}Recursive construction of FFT graph. 
$F_n$ is built from two copies of $F_{n-1}$ by adding an $(n+1)^{th}$ layer of $2^n$ 
vertices and connecting them with vertices of the $n^{th}$ layer using
butterflies.}
\end{figure}

Our construction is based on embedding an FFT butterfly graph into a 
triangulation of a three-dimensional cube. 
The $2^n$-point FFT graph, denoted by $F_n$, has $(n+1)2^n=N$ 
vertices, arranged in $n+1$ layers of $2^n$ vertices each, see Figure 
\ref{fig:FFT_NETWORK_FIG}. 
In other words, vertices of $F_n$ form an array $(k,i), ~ 0\leq k\leq n, 0\leq 
i\leq 2^n -1$, and a vertex $(k,i), ~k<n$ is connected with vertices $(k+1,i)$ 
and $(k+1,i \bigoplus 2^k)$, where $i \bigoplus 2^k$ signifies taking the 
complement of the $k^{th}$ bit of $i$. 

\begin{theorem}\label{th:FFT_LOWER_BOUND_POINTWISE} 
The number of cache misses $\mu$ in the evaluation of a symmetric, first-order
operator on $F_n$ is bounded as follows:
\begin{equation}\nonumber
\mu \geq \frac{1}{\cacheb}N(1+ \frac{\const}{\log S}) \, ,
\end{equation}
where $\const$ is a constant and $N=(n+1)2^n$.
\end{theorem}

\begin{proof}
The theorem is a direct sequence of Theorem \ref{th:LOWER_BOUND_POINTWISE} of 
Section \ref{sec:LOWER}, and of an estimation of the size of the boundary of 
vertex coverings of $F_n$, given in Proposition \ref{prop:FFT_SUM_OF_BOUNDARIES}.
\end{proof}

\begin{proposition}\label{prop:FFT_SUM_OF_BOUNDARIES}
Let $F_n= \bigcup V_i,~i=1,\dots ,k,~|V_i| \leq S$ be any partition, and 
$S\leq 2^{n/24}$.
Then the following inequality holds for the sum of the numbers of boundary
vertices of the sets of the partition:
\begin{equation}\label{eq:FFT_ISOPER_INEQ}
\sum_{i=1}^{k} |\partial(V_i)| \geq {N \over {4\log S}} \, .
\end{equation}
\end{proposition}
\begin{proof}
For any subset $V \subset F_n$ it follows from (\ref{eq:FFT_ISOPER_INEQ0}) that 
$\delta(V) \geq {1 \over 2} |V| /\log|V|$, and we can estimate the sum of 
boundaries of the partition as follows: 
\begin{equation}\nonumber
\begin{aligned}
\sum_{i=1}^{k} |\partial(V_i)| \geq \sum_{i=1}^{k} \delta(V_i) -6 \cdot 2^n & 
\geq {1 \over 2} \sum_{i=1}^{k} {{|V_i|} \over {\log |V_i|}} 6 \cdot 2^n 
\nonumber \\
& \geq {1\over {2 \log S}} \sum_{i=1}^{k} |V_i| -6 \cdot 2^n \geq {N \over 
{4\log S}}\, , \nonumber 
\end{aligned}
\end{equation}
where the boundary operator $\delta$ is as defined in Lemma 
\ref{lm:FFT_RIGHT_SURFACE_TO_VOLUME}.
The last inequality holds since $S\leq 2^{n/24}$.
\end{proof}

\begin{lemma}\label{lm:FFT_RIGHT_SURFACE_TO_VOLUME}
Let $V$ be any nonempty node subset of grid $F_n$, $n \geq 1$, and let $E$ be 
the set of all edges of $F_n$ incident on at least one node of $V$.
Then we have 
\begin{equation}\label{eq:FFT_ISOPER_INEQ0}
|V| \leq 2 \delta(V) \log \delta(V) \, ,
\end{equation}
where $\delta(V)$ is sum of the number of boundary nodes in $V$ and of the number 
of boundary edges of $E$. A node of $V$  is called a boundary node if it is in 
layer 0 or $n$ of $F_n$, and an edge of $E$ is called a boundary edge if it 
connects a vertex in $V$ with a vertex not in $V$.
\end{lemma}
\begin{proof}
Our proof of inequality (\ref{eq:FFT_ISOPER_INEQ0}) is based on induction and is 
similar to the proof of Theorem 4.1 in \cite{PEBBLE}. 
The base of induction for $n=1$ is obviously true.
Next, let $V$ be partitioned into three sets $A$, $B$ and $C$, as shown in Figure 
\ref{fig:FFT_ISOP_STEP_FIG}.
Sets $A$ and $B$ contain the nodes of $V$ that are in the two separate
$F_{n-1}$ subgraphs of $F_n$.
Sets $A_0$ and $B_0$ are contained in the last layer of these $F_{n-1}$.
Set $C$ contains the $n^{th}$ layer of nodes of $V$ in $F_n$.
The subset of nodes of $F_n$ that constitutes $V$ is indicated by open circles.
\begin{figure}[htb]
\centerline{\epsfig{file=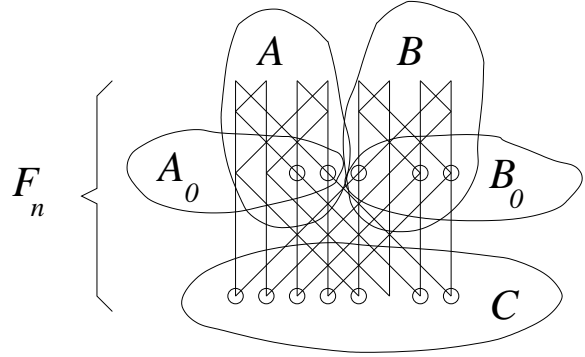,width=3in}}
\caption{\label{fig:FFT_ISOP_STEP_FIG} Induction step for proving the 
surface-to-volume inequality of a subset in $F_n$. We can assume without loss of 
generality that $|A_0| \geq |B_0|$.}
\end{figure}

Further, we define the partition $C=C^2 \cup C^1 \cup \ C^0$, where $C^i$ is the
subset of nodes in $C$ that are adjacent to exactly $i$ nodes in $A \cup B$. 
Now the equality
\begin{equation}\nonumber
|V| = |A|+|B|+|C^2|+|C^1|+|C^0|
\end{equation}
follows from the fact that $A\cup B\cup C$ is a partition of $V$, with $A$ 
and $B$ contained in $F_{n-1}$.

Now consider those edges of the last layer of $F_n$ that are incident on a 
vertex of $V$. 
These edges can be partitioned into 5 groups: those incident on nodes in $A$ 
and $C$, $B$ and $C$, $A$ only, $B$ only, and $C$ only.
We denote these sets by $E_{AC}, E_{BC}, E_A, E_B$ and $E_C$, respectively.
Since two such edges are incident on each vertex in $A$, $B$ and $C$,
we have the following two relations:

\begin{equation} \nonumber
|E_A|+|E_{AC}|+|E_B|+|E_{BC}|=2(|A_0|+|B_0|)
\end{equation}
\begin{equation} \nonumber
|E_C|+|E_{AC}|+|E_{BC}|=2|C|
\end{equation}
From these equalities it follows that
\begin{equation} \nonumber
|E_A|+E_B|+|E_C|=2(|A_0|+|B_0|-|C|)+2|E_C|.
\end{equation}

We determine the size of the boundary of $V$ by adding to $A \cup B$
the nodes contained in the last graph layer of $F_n$, i.e.\ those in
$C$.
This causes some boundary nodes of $A$ and $B$ to cease to be boundary
nodes, namely those in $A_0$ and $B_0$, while all nodes in $C$ are
now new boundary nodes of $V$.
Subsequently, we determine the effect of adding $C$
on the number of boundary edges.
It is easy to see that all edges in $V$ that were boundary edges of $A$
or $B$ are also boundary edges of $V$.
The boundary edges that are newly created by adding $C$ are those that
connect a node of $A_0$ or $B_0$ to a node not in $C$, as well as
those that connect a node of $C$ to a node not in $A_0$ or $B_0$.
These new edge sets are exactly the following: $E_A$,  $E_B$, and  $E_C$,
which are all disjoint.
Consequently, we find:
\begin{equation}
\delta(V)=\delta(A)+\delta(B)-|A_0|-|B_0|+|C|+|E_A|+|E_B|+|E_C|.
\end{equation}

Taking into account that $|C^2|\leq 2|B_0|$ and that $|E_C|=|C^1|+2|C^0|$, from 
the previous equations we derive
\begin{equation} \label{eq:FFT_BOUNDARY_RELATION}
\delta(V)\geq \delta(A)+\delta(B)+|A_0|-|B_0|+|C^1|+3|C^0| \, .
\end{equation}

Using induction, and assuming that $\delta(B)$ is nonzero, we obtain
\begin{equation}\nonumber
|V|\leq 2(\delta(A) \log \delta(A) +\delta(B) \log \delta(B)+ |B_0|) =
2(\delta(V) \log \delta(V) - X)
\end{equation}
\noindent where 
\begin{equation}\nonumber
\begin{aligned}
X & \geq(\delta(A) + \delta(B) + D + |C^1| + 3|C^0|)
    \log (\delta(A) + \delta(B) + D + |C^1| + 3|C^0|)
 \nonumber \\ 
  &-\delta(A) \log \delta(A) - \delta(B) \log \delta(B) - |B_0| \, ,\nonumber 
\end{aligned}
\end{equation}
\noindent  and $D=|A_0|-|B_0|$.
Because $D \geq 0$, $|C^1| + 3|C^0| \geq 0$, and $\delta(A), \delta(B) \geq 1$,
we get
\begin{equation}\nonumber
X  \geq \delta(A) \log (1+{{\delta(B) } \over {\delta(A)}} ) 
  + \delta(B) \log (1+ {{\delta(A)} \over {\delta(B)}} ) -|B_0| \, .
\end{equation}
Since $|B_0| \leq \delta(B)$ and $|B_0| \leq \delta(A)$ we finally obtain:
\begin{equation}
X \geq |B_0| \log \left[ 1 + 
             \left( \frac{\delta(A)}{\delta(B)} +
                    \frac{\delta(B)}{\delta(A)} \right)/2 \right]
  \geq 0 \, .
\end{equation}
If $\delta(B)$ is zero, the proof simplifies significantly, since now we
have to show nonnegativity of $X$, with
\begin{equation}
X \geq (\delta(A) + |A_0| + |C^1| + 3|C^0|) 
       \log (\delta(A) + |A_0| + |C^1| + 3|C^0|) - \delta(A) \log \delta(A) \, ,
\end{equation}
which is trivial, since $\delta(A) \geq 1$.
\end{proof}

\begin{figure}[hbt]
\begin{minipage}[t]{2.25in}
\centerline{\epsfig{figure=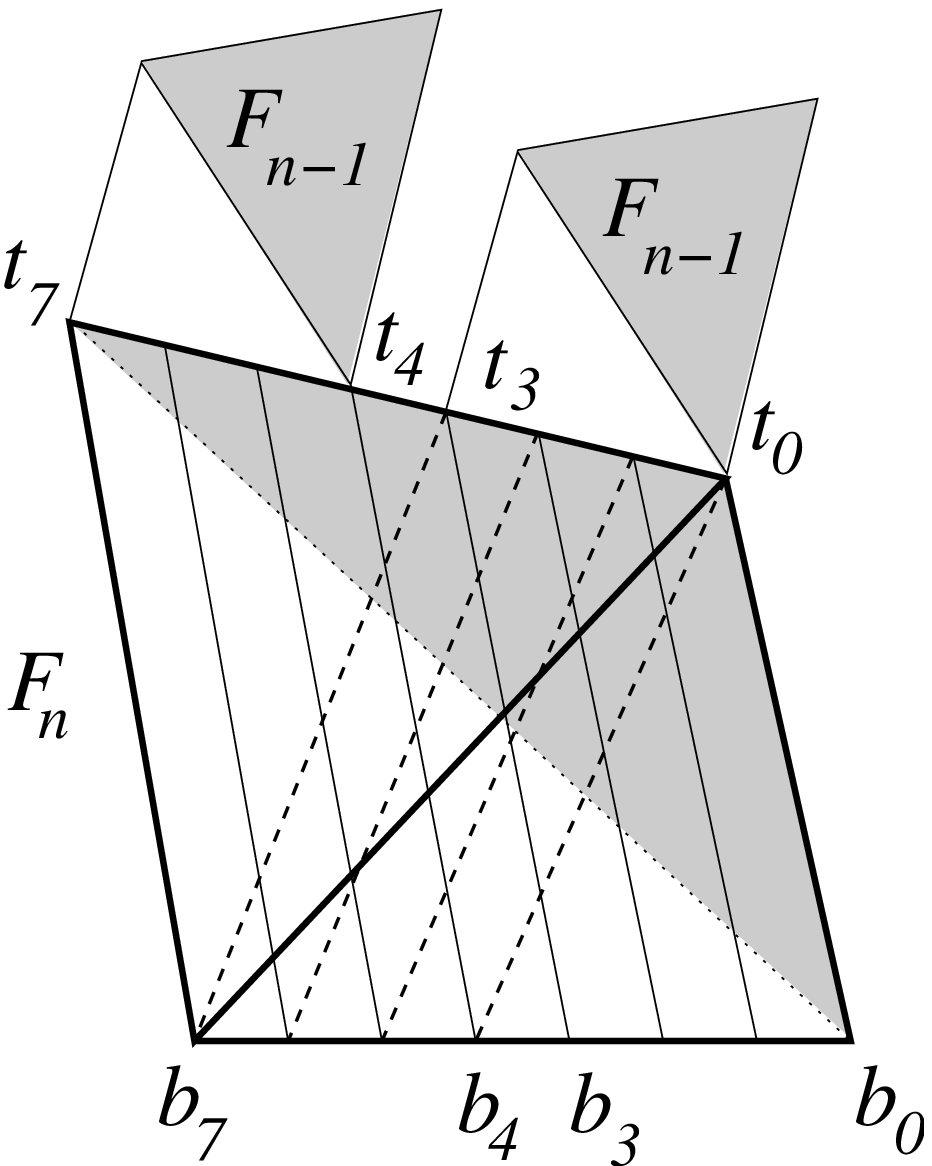,height=2.25truein}}
\caption{\sf \noindent \label{fig:FFT_TRIANG_FIG} Recursive construction of embedding 
of FFT graph into a triangulations of a simplex.}
\end{minipage}
\hspace{.2in}  % allow some horizontal space between the figures
\begin{minipage}[t]{2.25in}
\centerline{\epsfig{figure=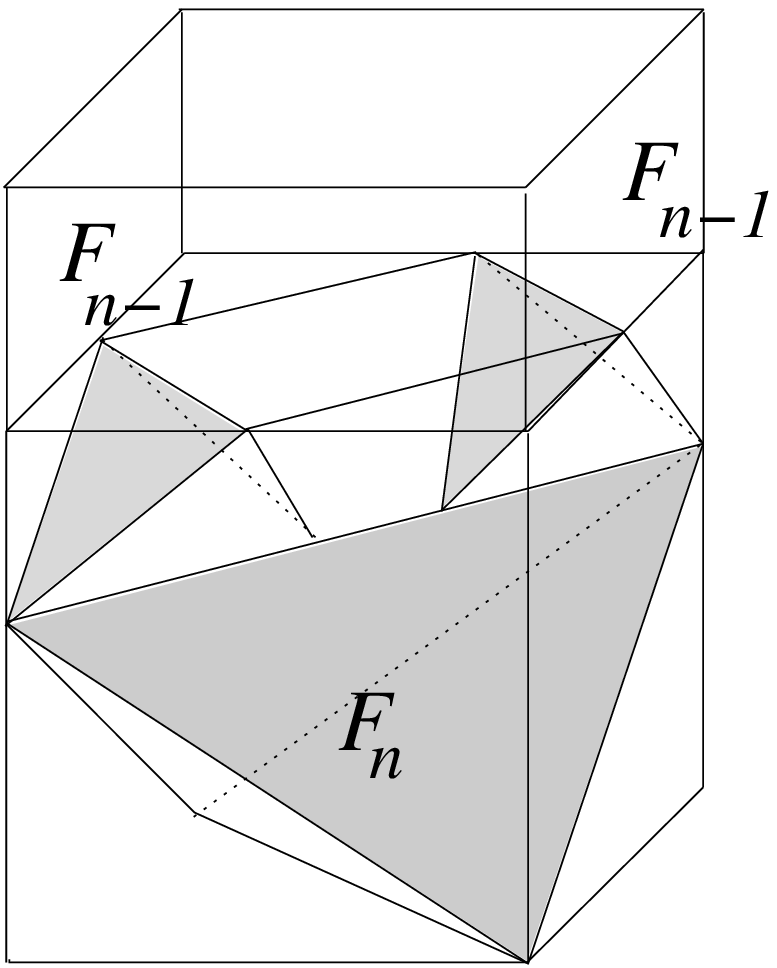,height=1.6truein}}
\caption{\sf \label{fig:CUBE_TRIANG_FIG} (Partial) recursive triangulation of a cube 
(view rotated with respect to Fig.\ \ref{fig:FFT_TRIANG_FIG}).}
\end{minipage}
\end{figure}

The FFT graph can be embedded into a triangulation of a three-dimensional cube. 
An inductive step of embedding of the FFT graph into a triangulation of simplices 
is shown in Figure \ref{fig:FFT_TRIANG_FIG}. 
The simplices can be embedded into a cube, as shown in Figure 
\ref{fig:CUBE_TRIANG_FIG}, which then can be partitioned into parallelepipeds.
Finally, each such parallelepiped can be triangulated.

\begin{figure}[hbt]
\begin{minipage}[t]{2.25in} 
\centerline{\epsfig{figure=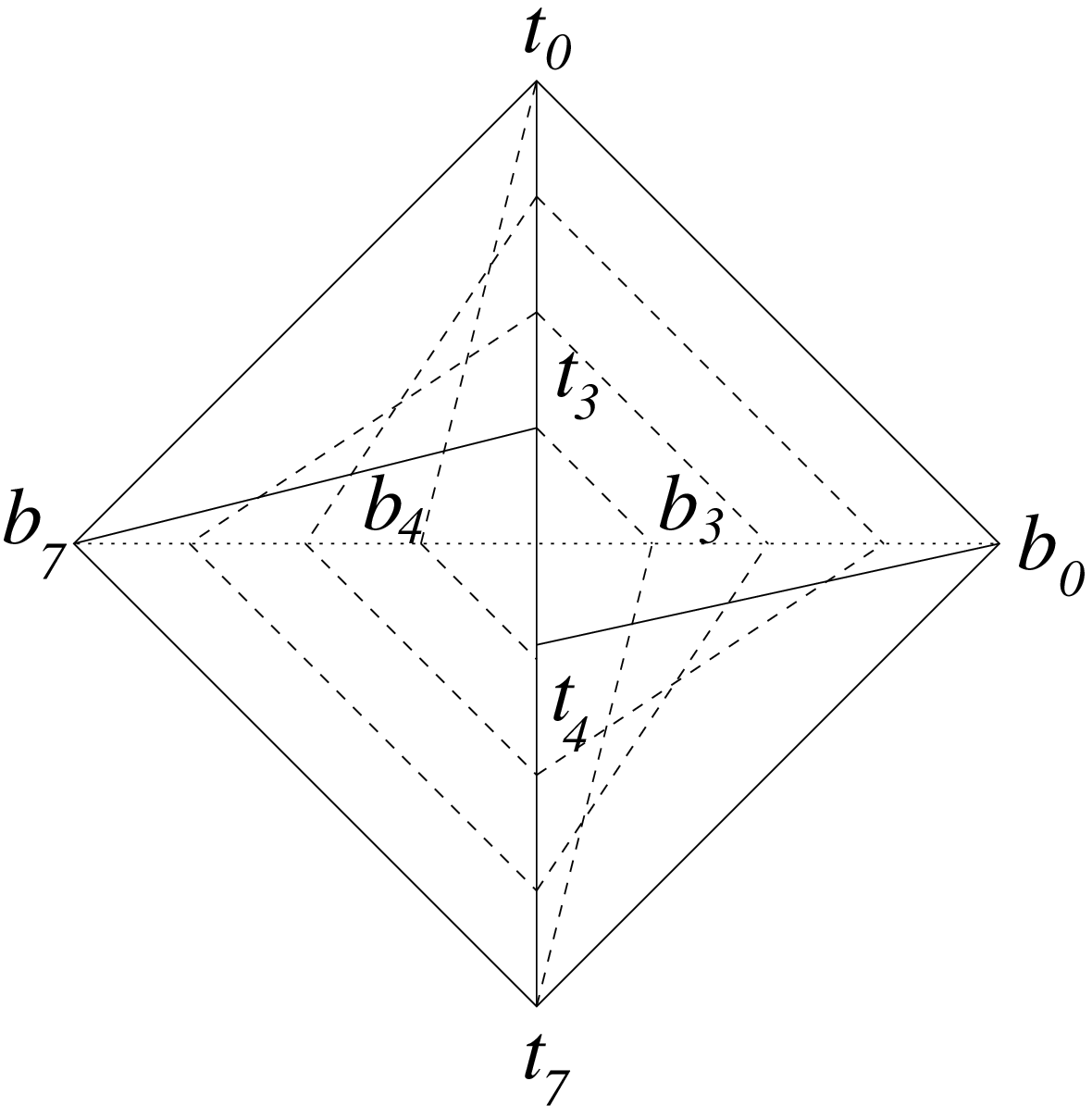,height=2.25truein}}
\noindent
\caption{\sf \label{fig:TRIANGILATION_TOP_FIG} Embedding one layer of an FFT graph 
         into a triangulation of a simplex, top view.}
\end{minipage}
\hspace{.2in}
\begin{minipage}[t]{2.25in}
\centerline{\epsfig{figure=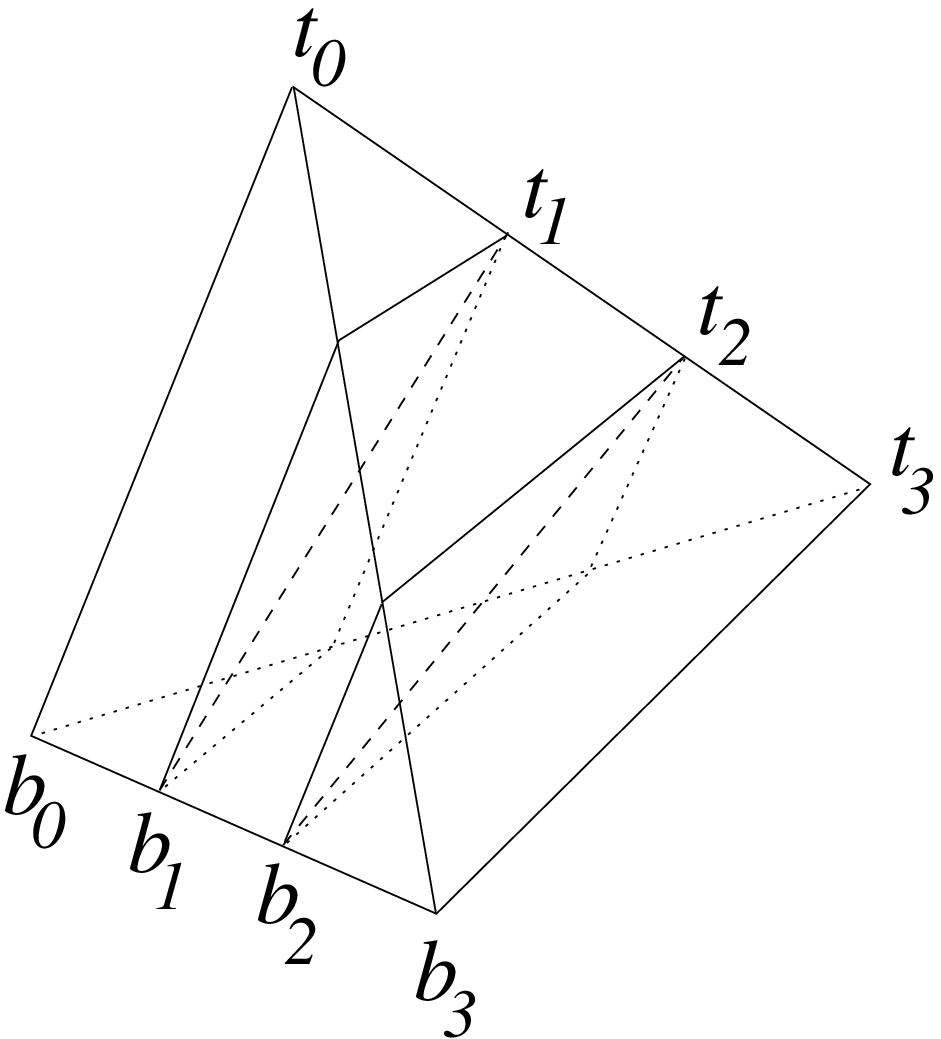,height=1.6truein}}
\caption{\sf \label{fig:rulered_SURFACE_FIG} Triangulation of a simplex separated 
by a ruled surface via plane sections parallel to the edges $(b_0,t_0)$ and $(b_3,t_3)$.
%$s_1,s_2,s_3$. Only partition not shadowed by the ruled surface is shown.
}
\end{minipage}
\end{figure}

Assuming that we have an embedding of the edges connecting the last two layers 
of $F_{n-1}$ (butterflies) into the triangulation of a simplex, we embed the 
butterflies connecting the last two layers of $F_n$ into a triangulation of a 
simplex.
We map the vertices of the last two layers $F_n$ into equidistant points on
two crossing edges of a simplex, see Figure \ref{fig:FFT_TRIANG_FIG}. 
If we linearly parametrize these crossing edges and connect corresponding points 
by lines, we build a ruled surface. 
A ruled surface can be viewed as a hyperboloid containing the two crossing edges 
and the connecting lines.
Each ruled surface separates a simplex built on the appropriate vertices 
into two parts, as shown in Figure \ref{fig:rulered_SURFACE_FIG}; 
the top view is shown in Figure \ref{fig:TRIANGILATION_TOP_FIG}. 

We map the edges of the butterflies onto lines of one of the ruled surfaces 
separating simplices $(t_0,t_3,b_7,b_4)$ and $(t_7,t_4,b_3,b_0)$ and two 
other ruled surfaces separating simplices $(t_0,t_3,b_3,b_0)$ and 
$(t_7,t_4,b_4,b_7)$, respectively.
The whole simplex  $(t_0,t_7,b_7,b_0)$ can be partitioned into the four 
simplices listed above,  and 5 primitive simplices which will not be further 
partitioned: $(t_3,t_4,b_3,b_4)$, $(t_3,t_4,b_4,b_7)$, $(t_3,t_4,b_3,b_0)$, 
$(t_0,t_3,b_4,b_3)$ and $(t_7,t_4,b_4,b_3)$. 
Each of the simplices $(t_0,t_3,b_7,b_4)$, $(t_7,t_4,b_3,b_0)$, 
$(t_0,t_3,b_3,b_0)$ and $(t_7,t_4,b_4,b_7)$ is divided by a ruled surface, 
hence it is sufficient to build a triangulation of a simplex separated by a 
ruled surface. 
This can be done in 2 steps: 
1. partitioning the simplex into prisms by planes parallel to the other 
   crossing edges of the simplex (see Figure \ref{fig:rulered_SURFACE_FIG}), 
   and 
2. triangulating each of the prisms into primitive simplices.

Finally, we embed the triangulated simplices into a cube and 
augment it by a triangulation of the space between simplices and the cube,
as shown in Figure \ref{fig:CUBE_TRIANG_FIG}.

It is easy to verify that the total number of vertices in the triangulation 
does not exceed $M=3n2^n$, and that the degree of each node does not exceed 16. 
Hence we have constructed a triangulation having the property declared at 
the beginning of this section.
\noindent 

\section{Related Work and Conclusions}\label{sec:CONCLUSIONS}

Because of the growing depth of memory hierarchies and the concomitant
increase in data acess times, the reduction of cache misses in scientific 
computations remains an active subject of research. 
One of the first lower bounds for data movement between primary and 
secondary storage was obtained in \cite{PEBBLE}. 
Recent work has focused on developing compiler techniques to 
reduce the number of cache misses. 
In this direction we mention \cite{MARTONOSI}, where the notion of  
cache miss equation (CME) and a tiling of structured grids with 
conflict-free rectilinear parallelepipeds were introduced. 
Some tight lower and upper bounds for computation of local operators
of fixed radius on structured grids were obtained in 
\cite{FRUMKIN_WIJNGAART}, where a tiling with a reduced fundamental 
parallelepiped of the interference lattice was used for reduction of 
cache misses. 
Some practical methods for improving cache performance in computations 
of local operators are given in \cite{RIVERA_SC00}.

In this paper we showed that the reduction of cache misses for computations 
of local operators of fixed radius, defined on structured or unstructured 
discretization grids, is closely related to the problem of covering these 
grids with conflict-free sets having low surface-to-volume ratio. 
We introduced two new coverings of structured grids: one with 
Voronoi cells, and one with rectilinear parallelepipeds built on 
the vectors of successive minima of the grid interference lattice. 
The cells of both coverings have near-minimal surface-to-volume ratios. 
Direct measurements of cache misses show a significant advantage of the 
successive minima covering relative to computations using the natural 
loop order, maximally optimized by a compiler. 
We also showed that the computation of local operators of fixed radius 
on planar unstructured grids can be organized in such a way that the 
number of cache misses is asymptotically close to that on structured grids.
Finally, we demonstrated how the latter result can be extended to
higher dimensions.

\section*{Appendix: The Weight Inequality}

\begin{proposition} \label{th:convex_roots_d}
For any positive $v_1 \geq \dots \geq v_k >0$ such that $v_1 \leq \sum_{i=2}^{k} v_i$, 
the following inequality holds:
\begin{equation}\nonumber
\sum_{i=1}^{k} v_i^{1/\grdim} \geq \left( 2\sum_{i=1}^{k} v_i \right) ^{1/\grdim}.
\end{equation}
\end{proposition}

\begin{proof}
The proof uses Jensen inequality, see \cite{HARDY}, Ch 2.10, Th. 19: for $0<r<s$
\begin{equation}\nonumber
(\sum_{i=1}^{k} v_i^{1/r} )^r \leq (\sum_{i=1}^{k} v_i^{1/s})^s
\end{equation}
\noindent and in particular
\begin{equation}\nonumber
(\sum_{i=1}^{k} v_i^\grdim )^{1/\grdim} \leq 
\sum_{i=1}^{k} v_i \leq (\sum_{i=1}^{k} v_i^{1/\grdim})^\grdim
\end{equation}
Let $x_i=v_i^{1/\grdim},~i=1,\dots ,k$, we have to prove
\begin{equation}\nonumber
(\sum_{i=1}^{k} x_i)^\grdim \geq 2 \sum_{i=1}^{k} x_i^\grdim
\end{equation}
where $x_1 \leq (\sum_{i=2}^{k} x_i^\grdim)^{1/\grdim} \leq \sum_{i=2}^{k} x_i$ and $x_1 \geq \dots \geq x_k>0$.

Let $j<k$ be the minimal index such that  $x_j > \sum_{i=j+1}^{k} x_i$. 
Obviously, $j>1$, so we have:
\begin{equation}\nonumber
\begin{aligned}
(\sum_{i=1}^{k} x_i)^\grdim> & \sum_{i=1}^{k} x_i^\grdim +\grdim x_1^{\grdim-1}\sum_{i=2}^{k} x_i+ \cdots +\grdim x_{j-1}^{\grdim-1} \sum_{i=j}^{k} x_i+\grdim x_j^{\grdim-1}\sum_{i=j+1}^{k} x_i \nonumber \\
> & \sum_{i=1}^{k} x_i^\grdim +\grdim x_1^\grdim+ \cdots +\grdim x_{j-1}^\grdim +\grdim x_j^{\grdim-1}(x_{j+1}+ \cdots + x_k) \nonumber \\
\geq & \sum_{i=1}^{k} x_i^\grdim +\grdim x_1^\grdim + \cdots +\grdim x_{j-1}^\grdim+\grdim x_{j+1}^\grdim+ \cdots + \grdim x_k^\grdim > 2 \sum_{i=1}^{k} x_i^\grdim \nonumber 
\end{aligned}
\end{equation}
since $\grdim \geq 2$. If no such $j$ exists then the proof becomes even simpler. 
\end{proof}

\end{document}